\documentclass[table,xcdraw,dvipsnames,acmsmall,screen]{acmart}

\usepackage{booktabs}
\usepackage{multirow}
\usepackage[most, skins]{tcolorbox}
\usepackage{pifont}
\usepackage[ruled,linesnumbered]{algorithm2e}
\usepackage{xspace}
\usepackage{enumitem}
\usepackage{arydshln}

\newcommand\dashline{\arrayrulecolor{gray}\hdashline[1pt/1pt]\arrayrulecolor{black}}

\newcommand{\artifacturl}{\url{https://github.com/pan2013e/catcoder}}

\newcommand{\appname}{\textsc{CatCoder}\xspace}

\newcommand{\vanilla}{\textsc{Vanilla}\xspace}
\newcommand{\infile}{\textsc{In-File}\xspace}
\newcommand{\repocoder}{\textsc{RepoCoder}\xspace}

\newcommand\myding[1]{\ding{\numexpr181+#1\relax}}
\newcommand{\dingcheckmark}{\ding{51}}

\newcommand{\rqans}[2]{
\begin{tcolorbox}[
    left=4pt, right=4pt, top=4pt, bottom=4pt,
    boxrule=0.2mm,
    leftrule=2mm,
    arc=0mm,
    colframe=black!40!white,
    colback=black!5!white,
    colbacktitle=black!50!white
]
\textbf{Answer to RQ#1:~}{#2}
\end{tcolorbox}
}

\definecolor{codeborder}{RGB}{240, 240, 240}
\NewTotalTCBox{\code}{ s v }
{verbatim,colback=white,colframe=codeborder,boxsep=0mm,left=1pt,right=1pt,top=1pt,bottom=1pt}
{\lstinline^#2^}

\begin{document}

\title{\textsc{CatCoder}: Repository-Level Code Generation with Relevant Code and Type Context}

\newcommand{\zju}{The State Key Laboratory of Blockchain and Data Security, Zhejiang University}
\author{Zhiyuan Pan}
\affiliation{
  \institution{\zju}
  \city{Hangzhou}
  \country{China}
}
\email{zy\_pan@zju.edu.cn}

\author{Xing Hu}
\authornote{Corresponding Author}
\affiliation{
  \institution{\zju}
  \city{Hangzhou}
  \country{China}
}
\email{xinghu@zju.edu.cn}
            
\author{Xin Xia}
\affiliation{
  \institution{\zju}
  \city{Hangzhou}
  \country{China}
}
\email{xin.xia@acm.org}

\author{Xiaohu Yang}
\affiliation{
  \institution{\zju}
  \city{Hangzhou}
  \country{China}
}
\email{yangxh@zju.edu.cn}

\received{08 January 2025}
\received[Revised]{03 August 2025}
\received[Accepted]{13 November 2025}

\begin{abstract}

Large language models (LLMs) have demonstrated remarkable capabilities in code generation tasks. However, \textit{repository-level code generation} presents unique challenges, particularly due to the need to utilize information spread across multiple files within a repository.
Specifically, successful generation depends on a solid grasp of both general, context-agnostic knowledge and specific, context-dependent knowledge. While LLMs are widely used for the context-agnostic aspect, existing retrieval-based approaches sometimes fall short as they are limited in obtaining a broader and deeper repository context.
In this paper, we present \textbf{\appname}, a novel code generation framework designed for statically typed programming languages. \appname enhances repository-level code generation by integrating relevant \underline{\textbf{c}}ode \underline{\textbf{a}}nd \underline{\textbf{t}}ype context. Specifically, it leverages static analyzers to extract type dependencies and merges this information with retrieved code to create comprehensive prompts for LLMs. To evaluate the effectiveness of \appname, we adapt and construct benchmarks that include 199 Java tasks and 90 Rust tasks. The results show that \appname outperforms the RepoCoder baseline by up to 14.44\% and 17.35\%, in terms of compile@$k$ and pass@$k$ scores. In addition, the generalizability of \appname is assessed using various LLMs, including both code-specialized models and general-purpose models. Our findings indicate consistent performance improvements across all models, which underlines the practicality of \appname. Furthermore, we evaluate the time consumption of \appname in a large open source repository, and the results demonstrate the scalability of \appname.
\end{abstract}

\ccsdesc[500]{Software and its engineering~Automatic programming}

\keywords{Large Language Model, Code Generation, Repository Context}

\maketitle

\section{Introduction}
\label{sec:intro}

Large language models (LLMs)~\cite{wang2021codet5, nijkamp2022codegen, codellama, starcoder2} have been widely applied in various software engineering-related fields, and they have shown remarkable performance in code generation tasks~\cite{humaneval, mbpp}.
Recently, repository-level code generation has attracted much attention~\cite{liu2023repobench,wu2024repoformer,zhang2023repocoder,shrivastava2023repository}. It refers to the generation of a specific function within a repository and is thus more practical than standalone code generation tasks (i.e., generating functions with no dependencies). This practical scenario introduces unique challenges, as the target code often depends on definitions and usage patterns distributed across multiple files, rather than being confined to a single file. When provided with only the \textit{in-file} context for such a task, an LLM may hallucinate and produce code that appears plausible but fails to compile or execute correctly within the project (e.g., misusing an API or referencing an undefined identifier). Therefore, a key problem in repository-level code generation is how to provide the LLM with the necessary \textit{cross-file} context so that it can generate valid and consistent code.

Existing studies have tried to tackle this problem by utilizing repository information with retrieval-based approaches. 
Zan et al.~\cite{private_library} train a dual-encoder model to retrieve API information from library documentation.
Lu et al.~\cite{lu2022reacc} build an external code database to retrieve similar code. 
Shrivastava et al. ~\cite{shrivastava2023repository} propose RLPG, in which they train a classifier to predict prompt proposals taken from repository files. 
Zhang et al.~\cite{zhang2023repocoder} propose RepoCoder, which iteratively retrieves similar code in a repository. 
However, these approaches still have limitations. First, methods that depend on external resources or require training specialized retrieval models often lack generalizability across unseen repositories or programming languages.
Second, the retrieved code may fail to capture all necessary field accesses and API calls required for correct generation. This issue is particularly pronounced in repositories with minimal code duplication~\cite{zhang2023repocoder}, or when the desired functionality is novel within the codebase. These limitations underscore the need for more comprehensive modeling of repository context.

In this paper, we aim to address the aforementioned limitations by capturing a broader repository context for the LLM.
We observe that, in addition to code retrieval, the types associated with a function can also serve as valuable references for LLMs. 
Specifically, these related types provide a rich set of fields and methods from which the LLM can select when generating code, thereby enhancing its understanding of repository-specific details.
We refer to this explicit information of types as \textit{type context}. 
The type context encapsulates localized knowledge embedded in nested scopes, serves as a complement to code retrieval, and helps LLMs develop a deeper understanding of the repository structure. Our work focuses on constructing type context for \textit{statically typed} programming languages\footnote{Statically typed programming languages require variable types to be declared at compile time, such as C, C++, Java, and Rust.}, where precise type information can be reliably extracted and leveraged. In contrast, dynamically typed languages (e.g., Python and JavaScript) determine variable types at runtime, making type information less accessible and more ambiguous. To extract type context, we utilize existing static analyzers.

Building on the concept of \textit{type context}, we propose a new repository-level code generation framework named \textbf{\appname}. By leveraging the combination of relevant \underline{\textbf{c}}ode \underline{\textbf{a}}nd \underline{\textbf{t}}ype context, \appname aims to improve the generation of code that is both logically sound and correctly integrated within its local environment. The framework operates in three stages.
First, it retrieves relevant code from the repository. 
Next, it invokes a static analyzer to obtain dependent types for the target function and constructs a textual representation of the type context. Finally, it integrates both the retrieved code and the type context into a single prompt and generates the target function using an LLM. \appname runs on top of a frozen LLM and does not require external databases or additional model training.

To evaluate the effectiveness of \appname, we adapt and construct repository-level code generation benchmarks that include 199 Java tasks and 90 Rust tasks in total. We compare \appname with vanilla LLM, the In-File context (i.e., contents within the same file), and RepoCoder~\cite{zhang2023repocoder}, a state-of-the-art framework for repository-level code generation, in terms of compile@$k$ and pass@$k$~\cite{humaneval} scores (i.e., compilation rate and test passing rate).
To validate the generalizability of \appname, we evaluate \appname with various LLMs, including code-specialized models and general-purpose models. To assess the scalability of \appname, we conduct extensive evaluations on a large open-source repository, measuring the time consumption of sequential code generation tasks. The results on the benchmarks indicate that \appname substantially outperforms the vanilla LLM and the In-File context. In addition, it outperforms RepoCoder by up to 14.44\% and 17.35\%, in terms of compile@$k$ and pass@$k$. The results of the generalizability study demonstrate that \appname improves the performance of all selected LLMs. Furthermore, the results of the scalability study underline the efficiency and practicality of \appname in large repositories.
 
\noindent\textbf{Contributions.} In summary, we make the following contributions:

\begin{enumerate}
    \item We propose a novel framework for statically-typed programming languages, \appname, to generate repository-level code by highlighting relevant \underline{\textbf{c}}ode \underline{\textbf{a}}nd \underline{\textbf{t}}ype context in code repositories.
    \item To the best of our knowledge, we are the first to construct a repository-level code generation benchmark for Rust, to evaluate the effectiveness of \appname on a minor programming language.
    \item We evaluate \appname using Java and Rust benchmarks, and results indicate that \appname outperforms the baselines.
    \item We release the replication package\footnote{\artifacturl} of \appname, including the source code of \appname and our benchmarks, to facilitate future research.
\end{enumerate}

\noindent\textbf{Paper Organization.} 
The remainder of the paper is structured as follows. Section~\ref{sec:motivation} introduces the motivation of \appname using an example. 
Section~\ref{sec:method} presents the details of \appname.
Section~\ref{sec:setup} describes the experimental setups, including baselines, evaluation metrics, benchmarks, and implementation details. 
Section~\ref{sec:results} analyzes the experimental results and provides answers to the research questions. 
Section~\ref{sec:discussion} discusses the failure cases of \appname and threats to validity.
Section~\ref{sec:related} summarizes the related work. Section~\ref{sec:conclusion} concludes the paper with possible future work.

\begin{figure}[t]
    \centering
    \includegraphics[width=\linewidth]{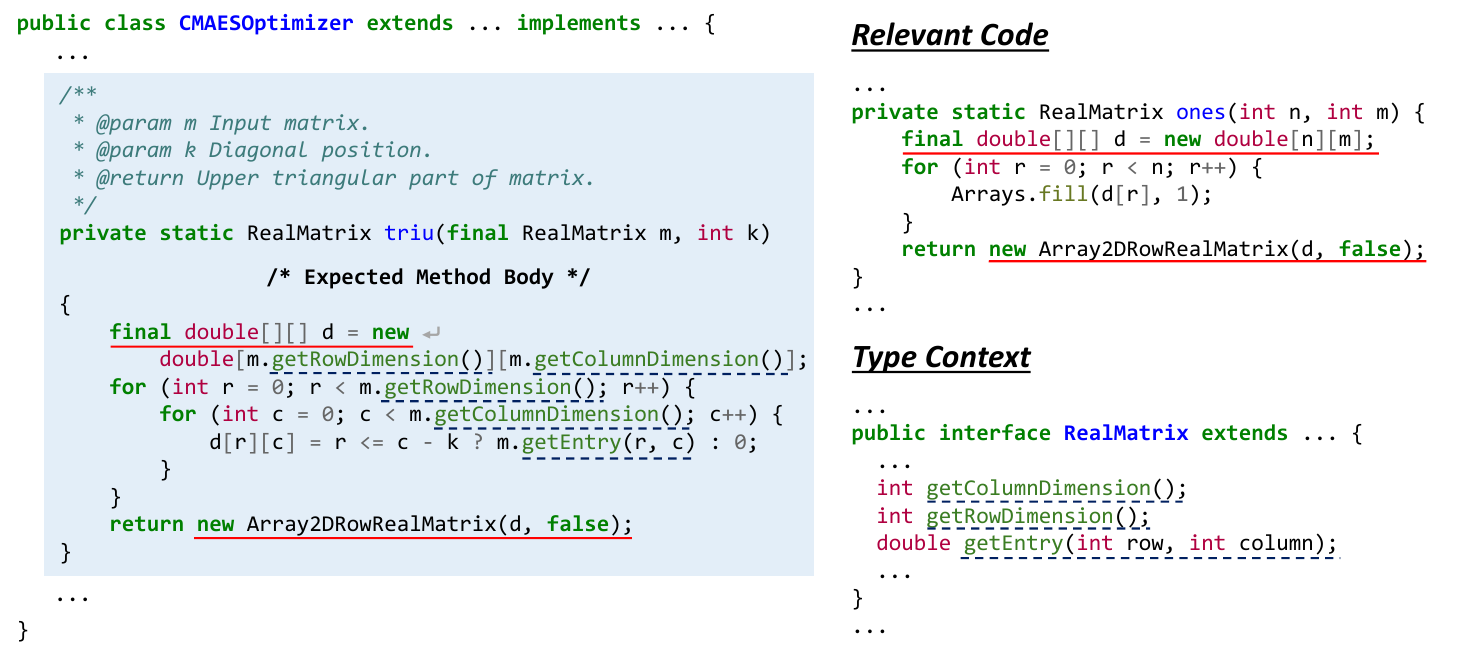}
    \caption{The two essential sources of local context for generating the \texttt{triu} method. (1) \textbf{Relevant Code} (the \texttt{ones} method) provides a crucial, repository-specific pattern for instantiating a \texttt{RealMatrix} object. (2) \textbf{Type Context} provides the equally crucial API definition for the \texttt{RealMatrix} interface, including methods like \texttt{getEntry} needed for element access.}
    \label{fig:scenario}
\end{figure}

\section{Motivating Example}
\label{sec:motivation}

In this section, we present a realistic development scenario to illustrate the core idea behind \appname. The motivating example will demonstrate that synthesizing information from both relevant code and type context is crucial for successful generation.

Consider the task of implementing the helper method \code{triu} within the \code{CMAESOptimizer} class from the widely used \textit{Apache Commons Math} library, as shown in Fig.~\ref{fig:scenario}. As described in its docstring, the \code{triu} method is intended to extract the upper triangular part of a given matrix. Its signature is \code{private static RealMatrix triu(final RealMatrix m, int k)}, where $m$ is the input matrix and $k$ specifies the diagonal offset.

Based on its context-agnostic knowledge of coding, an LLM can infer the general algorithmic principle: the implementation should return a new matrix with the same dimensions, where an element at position $(r, c)$ is copied from $m$ if the column index $c$ is greater than or equal to $r + k$; otherwise, the element is set to zero.
However, the requirement of context-dependent knowledge poses challenges to the LLM.
The \code{RealMatrix} type is an interface in \textit{Apache Commons Math} and cannot be treated as a nested \code{double} array. Proper interaction with a \code{RealMatrix} instance requires familiarity with repository-specific API conventions, such as using \code{getEntry(row, column)} to access elements. Furthermore, creating a new \code{RealMatrix} instance requires using a specific concrete implementation, \code{Array2DRowRealMatrix}, and understanding the correct constructor usage. An LLM must therefore grasp these repository-specific details to produce valid code.

\begin{table}[htbp]
\caption{Different information sources for the \texttt{triu} task}
\label{tab:motivation}
\begin{tabular}{@{}llc@{}}
\toprule
\textbf{Information Source}                           & \textbf{Key Insight Provided}    & \textbf{Sufficient Alone?} \\ \midrule
\multirow{2}{*}{\textbf{Info.1} \texttt{ones} method}   & Idiomatic instantiation:         & \multirow{2}{*}{\ding{56}} \\
                                                      & \texttt{new Array2DRowRealMatrix(...)}    &                            \\ \dashline
\multirow{3}{*}{\textbf{Info.2} \texttt{RealMatrix} API} & Correct API access:              & \multirow{3}{*}{\ding{56}} \\
                                                      & \texttt{m.getRow/ColumnDimension(...)}    &                            \\
                                                      & \texttt{m.getEntry(...)}                  &                            \\ \dashline
\appname (Info.1 + Info.2)                            & Complete, correct implementation & \ding{52}                  \\ \bottomrule
\end{tabular}
\end{table}

To generate a correct implementation, an LLM must solve two distinct subproblems, that is, (1) how to create the new \code{RealMatrix} to be returned, and (2) how to read data from the input \code{RealMatrix m}:

\begin{enumerate}
    \item As shown in Fig.~\ref{fig:scenario}, the \code{ones} method is retrieved due to its shared return type and location. From this snippet, an LLM can learn a non-trivial pattern for object creation: instantiate a new \code{Array2DRowRealMatrix} by first creating a primitive \code{double} array and passing it to the constructor. This form of relevant code retrieval reflects a common practice among developers, who often refer to similar methods when writing repository-level code.
    \item In addition, Fig.~\ref{fig:scenario} illustrates the available APIs of the \code{RealMatrix} type, presented as the type context. Because both the target method \code{triu} and the surrounding class \code{CMAESOptimizer} reference \code{RealMatrix}, its API information provides essential insights. For instance, an LLM can learn how to access dimensions and elements correctly by referencing the provided method signatures. This mirrors another common development strategy: navigating related types (e.g., parameter or return types) to discover usable fields and methods.
\end{enumerate}

Table~\ref{tab:motivation} categorizes the two aforementioned information sources and their key insights. We abstract them as follows: 

\begin{enumerate}
    \renewcommand{\labelenumi}{\textbf{Info.\theenumi}}
    \item Relevant code snippets that share similarities in method name, parameters, return types, or docstrings. These examples often perform semantically related or symmetric operations and thus provide reusable implementation patterns.
    \item Type context, defined as the available fields and methods of related types, offering clues for accessing or manipulating object instances correctly.
\end{enumerate}

This motivating example highlights that both \textbf{Info.1} and \textbf{Info.2} are indispensable. The absence of either is likely to lead to incorrect or incomplete code generation.
While prior work~\cite{lu2022reacc, zhang2023repocoder} has leveraged retrieval-augmented generation to utilize relevant code (Info.1), they typically overlook Info.2 (type context). As Fig.~\ref{fig:scenario} shows, retrieved code alone may not fully capture the required API usage. Moreover, in repositories that lack code duplication, relying solely on retrieval is less effective. In such cases, the absence of type context increases the risk of hallucination, such as referencing non-existent methods or fields. To address this gap, \appname integrates both Info.1 and Info.2. Our framework is carried out in the following steps:

\begin{enumerate}
\renewcommand{\theenumi}{\alph{enumi}}
\renewcommand{\labelenumi}{\textbf{Step \theenumi.}}

    \item \textbf{Relevant Code Retrieval.} We retrieve similar code in a repository based on the query (i.e., the docstrings and signature of the function to be completed).
    
    \item \textbf{Type Context Extraction.} We utilize static analyzers to determine the dependent types and then create type context by extracting the fields and method signatures.
    
    \item \textbf{Code Generation via LLM.} We merge the results in Step a and Step b into a comprehensive prompt, and invoke an LLM to obtain the generated code.
\end{enumerate}

\begin{figure}[t]
    \centering
    \includegraphics[width=\linewidth]{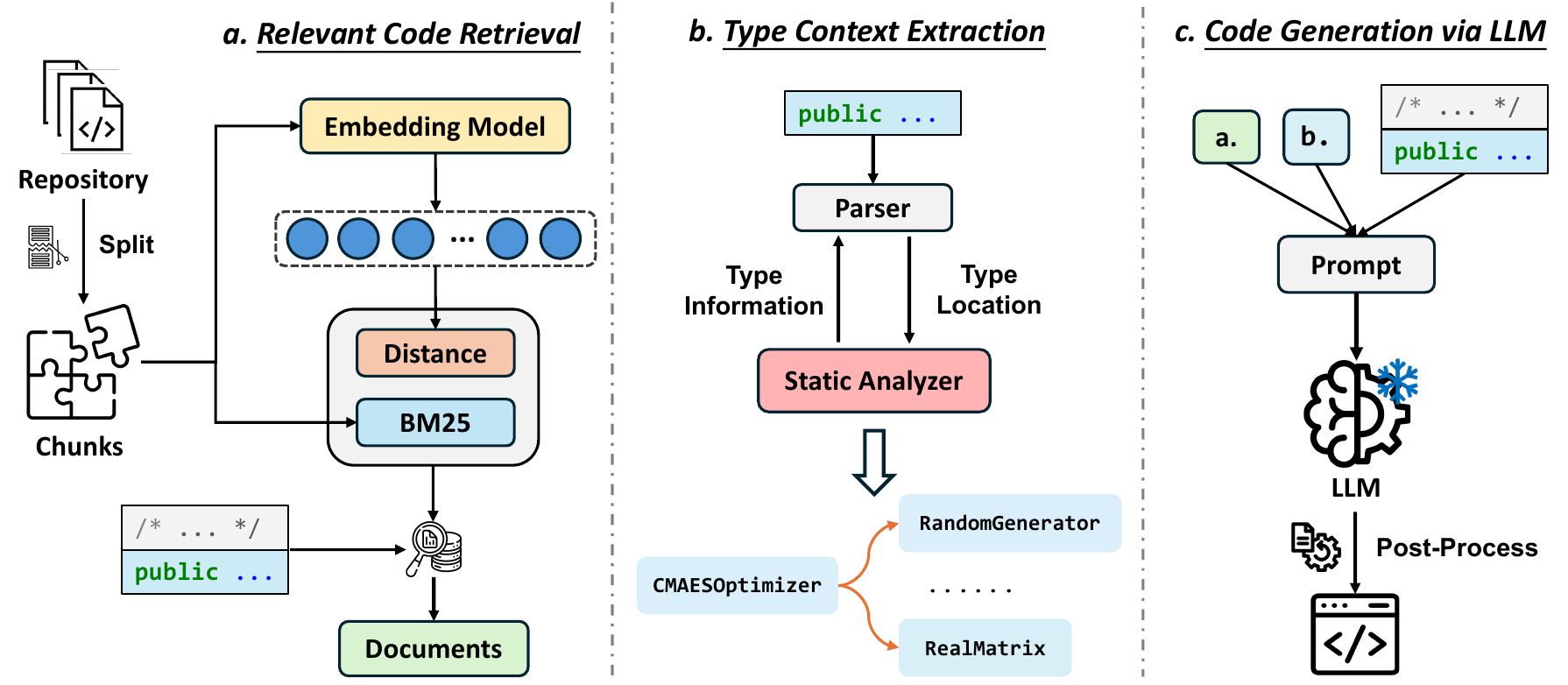}
    \caption{Overview of our approach}
    \label{fig:overview}
\end{figure}

\section{Proposed Approach}
\label{sec:method}

The framework of \appname is presented in Fig.~\ref{fig:overview}. 
\appname takes a source code repository and an empty function (along with its docstrings, if it exists). The output of \appname is the completed code of the function.

In addition to Fig.~\ref{fig:overview}, Algorithm~\ref{alg:overall} presents the pseudocode of the workflow of \appname. \appname contains two core components: (1) \textit{Relevant Code Retrieval} and (2) \textit{Type Context Extraction}. By utilizing these components, \appname is able to gather two kinds of referential information for the LLM. After post-processing the raw output of the LLM, \appname returns the generated code.

\SetKwComment{Comment}{/* }{ */}
\SetKwProg{Fn}{def}{:}{}
\begin{algorithm}[t]
    \DontPrintSemicolon
    \caption{\appname}
    \label{alg:overall}
    \KwIn{Code repository $R$, Empty function $f$}
    \KwOut{The completed code of function $f$}
    \BlankLine
    
    \SetKwFunction{FnRetrieval}{code\_retrieval}
    \SetKwFunction{FnContext}{type\_context}
    \SetKwFunction{FnPrompt}{prepare}
    \SetKwFunction{FnLLM}{LLM}
    \SetKwFunction{FnPost}{postprocess}
    
    \textit{docs} = \FnRetrieval{$R, f$} \Comment*[r]{Section~\ref{sec:retrieval}}
    \textit{contexts} = \FnContext{$R, f$} \Comment*[r]{Section~\ref{sec:context}}
    \textit{prompt} = \FnPrompt{$f$, docs, contexts} \Comment*[r]{Section~\ref{sec:codegen}}
    \textit{code} = \FnLLM{prompt}\;
    \KwRet{\FnPost{code}}\;
\end{algorithm}

\subsection{Relevant Code Retrieval}
\label{sec:retrieval}

\subsubsection{Code Splitting}

We traverse all the source code files in the repository and split them into smaller chunks. Starting with complete files, we set a maximum chunk size and recursively split larger fragments into smaller ones that fit in the chunk. 

As the structure of the source code is often complex, we cannot split it at an arbitrary position. For example, we should not break in the middle of a statement line or some important code elements (e.g., function definitions). Otherwise, the resulting contents would be less meaningful and difficult to understand. Therefore, we split the code only before the following positions (sorted by priority in descending order): 

\begin{enumerate}
    \item Type definitions (e.g., \code{public class Main} in Java)
    \item Function definitions (e.g., \code{public static void main(String[] args)} in Java)
    \item Control flow statements (e.g, \textit{if}, \textit{for} and \textit{while} statements in Java)
    \item New lines (i.e., \code{\n} and \code{\r\n})
\end{enumerate}

This rule follows the hierarchy of many programming languages and helps protect code structures during splitting. After splitting, we obtain a corpus of code chunks\footnote{We call them ``documents'' later in this subsection.} to perform retrieval.

\subsubsection{Hybrid Retrieving}

In the second step, we build retrievers to search for relevant code. To achieve better search performance, we follow previous research~\cite{luan2021sparse} and ensemble two kinds of retrievers: a \textit{sparse} one and a \textit{dense} one. The sparse retriever corresponds to keyword matching, and the dense retriever corresponds to semantic search. Each retriever searches and returns top-$k$\footnote{$k$ is a hyperparameter, and we describe the settings in Section~\ref{sec:exp_config}.} relevant documents (i.e., split code chunks), after a query is constructed by concatenating docstrings and the function signature.

\paragraph{\myding{1} Sparse Retrieval}

We choose the bag-of-words model for sparse retrieval, and use the BM25 function~\cite{robertson1995okapi, trotman2012towards} to rank the set of documents, based on the query terms.
Given a document $D$ and a query $Q$ with terms $t_1, t_2, \cdots, t_n$, the BM25 score can be defined as follows:

$$
\mathrm{BM25}(D, Q) = \sum_{t \in Q} \mathrm{IDF}(t,D) \cdot \dfrac{\mathrm{TF}(t,D) \cdot (k_1+1)}{\mathrm{TF}(t,D) + k_1 \cdot \left(1-b+b\cdot\dfrac{|D|}{|D|_{\mathrm{avg}}}\right)}
$$
where $\mathrm{IDF}(t, D)$ is the inverse document frequency weight, $\mathrm{TF}(t, D)$ is the term frequency (i.e., the number of appearing times in the document), $|D|$ is the length of the document, and $k_1$ and $b$ are hyperparameters. By default, the values of $k_1$ and $b$ are set to 1.5 and 0.75. $\mathrm{IDF}(t, D)$ can be calculated by:

$$
\mathrm{IDF}(t, D) = \log\dfrac{N_D -\mathrm{DF}(t, D) + 0.5}{\mathrm{DF}(t, D) + 0.5},
$$
where $N_D$ is the number of documents and $\mathrm{DF}(t, D)$ is the document frequency (i.e., the number of documents that contain the term).

In practice, after splitting the code files from a repository, we apply the aforementioned BM25 function for a query and obtain a list of scores. After that, we select documents with the top-$k$ highest scores.
 
\paragraph{\myding{2} Dense Retrieval}

We employ an embedding model for dense retrieval. The model maps natural language sentences and code snippets into dense vector spaces. To select the most relevant documents, our intuitive idea is to find the nearest neighbors around the embedding vector of the query text. Thus, we use \textit{Squared Euclidean Distance} to measure the proximity between embedding vectors in high-dimensional spaces, and this metric can be calculated by:

$$
d^2(\mathbf{p},\mathbf{q}) = \left\Vert \mathbf{p} - \mathbf{q} \right\Vert^2 = \sum_{i=1}^{n} (p_i - q_i)^2,
$$
where $n$ is the number of dimensions in the embedding vectors. A lower value of $d^2(\mathbf{p},\mathbf{q})$ indicates that $\mathbf{p}$ and $\mathbf{q}$ are more similar.

In practice, we utilize the embedding model and generate vector representations of all documents and query text. Then we calculate $d^2(\mathbf{p},\mathbf{q})$ between the query embedding and all document embeddings and select the top-$k$ nearest documents.

\subsubsection{Result Ranking}

In the third step, we ensemble the two retrievers and re-rank their outputs using the \textit{Reciprocal Rank Fusion} (RRF)~\cite{rrf} algorithm. We assign weights to the retrievers and calculate the weighted RRF score\footnote{We assume that a retriever does not output duplicate documents. However, duplicate documents across retrievers are acceptable.} for each retrieved document $D$:

$$
\mathrm{RRF}(D) = \sum_{i=1}^n \mathbf{1}(D)\cdot\dfrac{w_i}{r(D) + 60},
$$
where $w_i$ is the weight assigned to a certain retriever, $\mathbf{1}(D)$ is an indicator function\footnote{The value is 1 if $D$ is one of the outputs of this retriever, otherwise the value is 0.}, and $r(D)$ is the original rank (if exists) of $D$ in the outputs of this retriever.

After applying the RRF algorithm, we deduplicate the documents and sort them in descending order based on the RRF scores. Now we obtain the results of the \textit{Relevant Code Retrieval} component.

\subsection{Type Context Extraction}
\label{sec:context}

Here, we present the type context and the way to extract type contexts from code repositories using static analyzers\footnote{The word ``type'' is an abstract concept and might be different in various languages. For example, it refers to \textit{class}, \textit{interface}, and \textit{enum} in Java; whereas in Rust, it refers to \textit{struct} and \textit{enum}.}.

\subsubsection{Type Context}
\label{sec:type_context}

\begin{figure}[t]
\centering
\includegraphics[width=\linewidth]{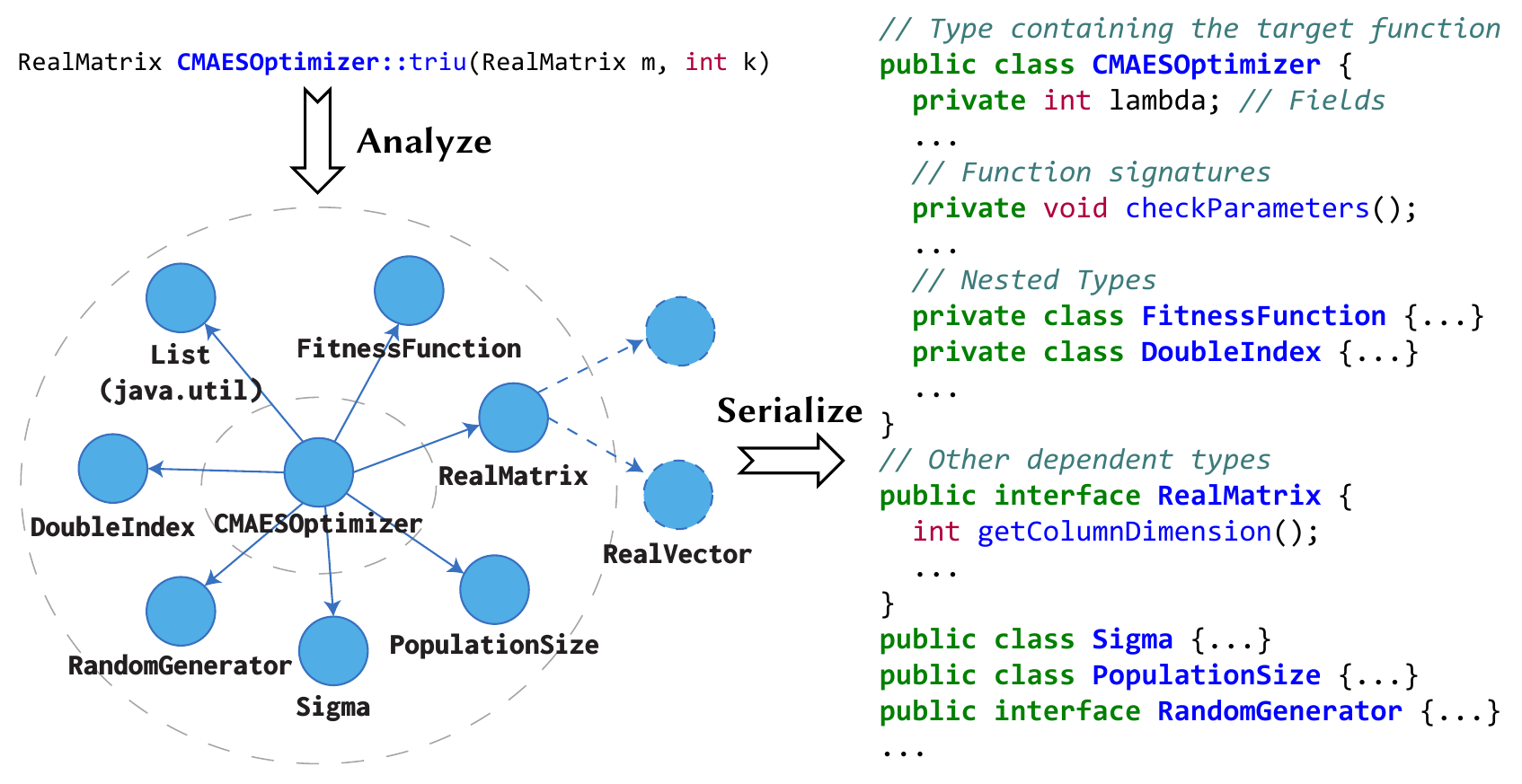}
\captionof{figure}{Dependency Graph and Type context of the \texttt{triu} method in class \texttt{CMAESOptimizer}.}
\label{fig:dep_graph}
\end{figure}

To capture the contextual relationships between types within a code repository, Fig.~\ref{fig:dep_graph} presents a directed dependency graph $G = (V, E)$, in which each vertex $v \in V$ corresponds to a type defined in the repository (e.g., a class or interface). A directed edge $(A, B) \in E$ is created from type $A$ to type $B$, if $A$ references or depends on $B$ (i.e., $B$ occurs in the body of $A$). The dependency graph is able to identify clusters of contextually related types and provide types that an LLM is likely to use and refer to when generating function code. 

In Fig.~\ref{fig:dep_graph}, we build the dependency graph for the \code{triu} method in class \code{CMAESOptimizer}, starting from its own class. As the graph spans, the vertex set and the edge set will eventually converge, and we can obtain the types that are relevant to \code{CMAESOptimizer}. To pass this structural information to the LLM, we create \textbf{type context} from the graph by extracting the fields and function signatures within the types and concatenating them into a string. Specifically, for the string representation, we follow the format proposed by Tufano et al.~\cite{tufano2020unit}, which has been applied in the field of unit test case generation. The right part of Fig.~\ref{fig:dep_graph} shows the resulting textual type context for the target method (simplified for illustration).

In large repositories, the full dependency graph can become vast, making the resulting type context too long to fit in an LLM's context window. However, the relevance between types tends to decrease with the distance between them in the graph. For example, in Fig.~\ref{fig:dep_graph}, programmers working on class \code{CMAESOptimizer} are more likely to directly use the API of \code{RealMatrix}, a direct neighbor, than the API of \code{RealVector}, which is an indirect neighbor at a distance of 2. Therefore, in practice, we constrain the scope of type context creation to the direct neighbors of the initial types in the graph. We also prune types from standard libraries, as we assume LLMs possess general knowledge of these libraries from their pre-training.

\subsubsection{Building Type Context via Static Analyzer}

To get detailed information (i.e., fields and methods) of a type from its name, some straightforward heuristics might be used, e.g., locating imports or looking for similar file names. However, these heuristics are not sound and do not work with all statically typed programming languages. Thus, we need to perform an analysis to accurately navigate to a type and find its definitions and implementations. Many language servers serve as static analyzers and provide this type-based analysis, such as clangd~\cite{clangd} for C++, Eclipse JDT~\cite{eclipsejdtls} for Java, and rust-analyzer~\cite{rust_analyzer} for Rust. 

Specifically, to build the type context for a function code generation task, we first determine the set of relevant types (as Section~\ref{sec:type_context} describes). We start with the signature and form a set of types, and the initial element is the type to which the current function belongs. Then we implement a parser that works collaboratively with the static analyzer. 
The static analyzer provides detailed type information to the parser. The parser then searches for a new occurring type, adds it to the set, and queries the analyzer with its location (i.e., byte offset or line and column offsets in a file). This interactive process continues until all direct neighbors of the initial set of types have been discovered. Finally, as described in Section~\ref{sec:type_context}, we concatenate the fields and method signatures within the set of types and obtain the type context.

\subsection{Code Generation via LLM}
\label{sec:codegen}

\begin{figure}[t]
    \centering
    \includegraphics[width=\linewidth]{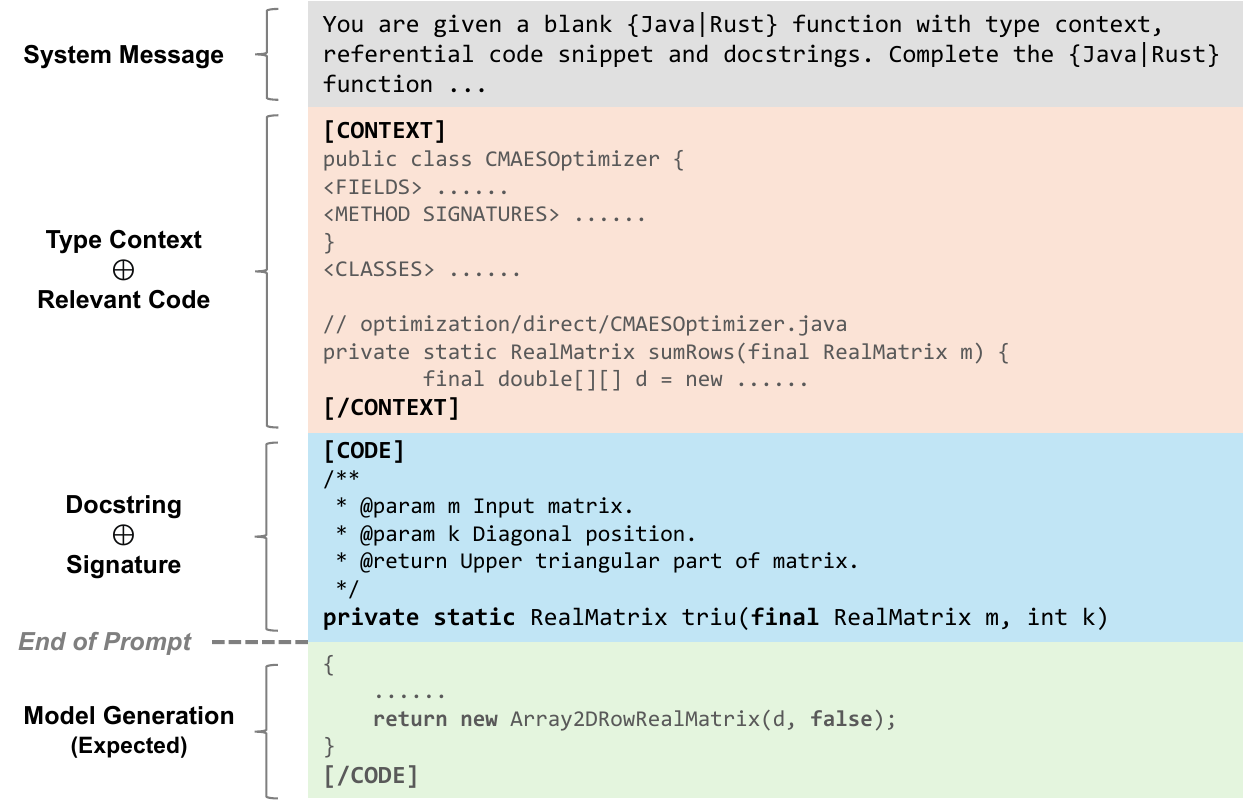}
    \caption{Our prompt design. The actual prompt ends right after the signature.}
    \label{fig:prompt}
\end{figure}

Finally, we describe how to utilize the previous results and invoke the LLM to generate code. Fig.~\ref{fig:prompt} illustrates our prompt design, in which there are two fields: \code{CONTEXT} and \code{CODE}. To construct a prompt, we follow the context format in previous work~\cite{zhang2023repocoder,shrivastava2023repository,liu2023repobench}, and fill the \code{CONTEXT} field by concatenating the type context and the retrieved code chunks. Then we place the docstrings and the signature of the method at the beginning of the \code{CODE} field. The prompt serves as input to a frozen LLM\footnote{We discuss the details of selected models and model access in Section~\ref{sec:impl}.}, and we expect the model to complete the \code{CODE} field by generating the method body (i.e., the \textcolor{ForestGreen}{\textbf{green}} part in Fig.~\ref{fig:prompt}). In addition, to post-process the raw output of the LLM, we employ several procedures\footnote{The post-processing procedures may vary with different LLMs and programming languages.}, including truncating texts, prepending function signatures, and removing Markdown syntax.

\section{Experimental Setup}
\label{sec:setup}

In this section, we first formulate four research questions (RQs). Based on the RQs, we describe the details of the baselines, evaluation metrics, benchmarks, and implementation of our approach.

We aim to answer the following research questions:

\begin{enumerate}
\renewcommand{\labelenumi}{\textbf{RQ\theenumi}}
    \item How effective is our approach at generating function code in a repository, compared with the baselines? And does our approach generalize to different LLMs?
    \item How do the two components in our approach (i.e., relevant code retrieval and type context extraction) contribute to the overall effectiveness, respectively?
    \item Is our approach generalizable to various large language models? And how do different models perform on the benchmark?
    \item Is our approach efficient when generating a large number of functions in large repositories?
\end{enumerate} 

In brief, we evaluate the \textbf{effectiveness} of our approach in RQ1, and then conduct an \textbf{ablation study} in RQ2. In RQ3, we study the \textbf{generalizability} of \appname and the performance of different models. Finally, in RQ4, we evaluate the \textbf{scalability} of our approach. For implementation and evaluation of effectiveness, we select two statically typed programming languages, Java and Rust, since Java is one of the most widely used programming languages~\cite{tiobe}, and Rust is a new system-level language that attracts the attention of many developers~\cite{rustlang}. 

\subsection{Baselines}

As described in Section~\ref{sec:codegen}, Fig.~\ref{fig:prompt} shows the prompt template used in our evaluation. As the \code{CONTEXT} field is the only variable, we compare our approach with methods that fill different contexts in the prompt. For the repository-level code generation task, there are three types of context: (1) \textit{No context}, (2) \textit{Local context}, and (3) \textit{Cross-file context}. Therefore, we design the following baselines for comparison in RQ1:

\begin{enumerate}
    \item \textbf{\vanilla}: In this method, we prompt LLMs without any repository context. In this case, all natural language descriptions are given, but the \code{CONTEXT} field in Fig.~\ref{fig:prompt} is left empty. 
    \item \textbf{\infile}: In this method, we consider code generation as a left-to-right file completion task. We extract code before the function to be generated and put it in the \code{CONTEXT} field.
    \item \textbf{\repocoder}: RepoCoder~\cite{zhang2023repocoder} is an approach that augments the prompt by iteratively retrieving referential code fragments, and it is originally proposed for Python code repositories. We adopt their idea and implement this baseline for Java and Rust based on our retrieval pipeline. Specifically, the initial iteration uses natural language descriptions (i.e., docstrings) as the query for code retrieval. From the second iteration, it uses the previously generated code for retrieval.
\end{enumerate}

\subsection{Evaluation Metrics}
\label{sec:metrics}

For the function code generation task, we argue that metrics based on text similarity are not accurate, and it is essential to use metrics based on compilation and execution results. Therefore, we use compilation rate and test passing rate, which evaluate syntax correctness and functional correctness, respectively. Specifically, to calculate such a rate, we follow Chen et al.~\cite{humaneval} and use the unbiased \textit{score}@$k$. It is defined as

$$
\text{score@$k$} = \mathop{\mathbb{E}}_{\text{Tasks}} \left[ 1 - \dfrac{{\binom{n-c}{k}}} {\binom{n}{k}} \right],
$$
where $n$ is the number of code samples generated per task and $k$ is the parameter in ``score@$k$'' ($n \ge k$). Based on this general formula, we define two detailed metrics that correspond to compilation rate and test passing rate:

\begin{itemize}
    \item \textbf{compile@\textit{k}}. In the formula, $c$ refers to the number of syntactically correct code samples. A code sample is syntactically correct if and only if it passes compiler checks and compiles.
    \item \textbf{pass@\textit{k}}. In the formula, $c$ refers to the number of functionally correct code samples. A code sample is functionally correct if and only if it passes the tests in the benchmark.
\end{itemize}

We calculate score@$k$ values for each task in the benchmark, and the final score@$k$ for a benchmark is the mean value of all score@$k$ values.

\subsection{Benchmarks}
\label{sec:benchmarks}

\subsubsection{Java Benchmark}

There are several publicly available repository-level benchmarks for Java~\cite{ding2024crosscodeeval, yu2024codereval, agrawal2024monitor, zeng2024coderujb}. However, it is challenging to use many of them to evaluate \appname because they either lack the raw repository code or are not designed for function code generation tasks. 
Considering the challenges, the CoderUJB~\cite{zeng2024coderujb} benchmark is a suitable base benchmark that contains 238 code generation tasks, and we successfully adapt it for our evaluation. In the adaptation process, we filter out a few functions that satisfy one of the following patterns:

\begin{enumerate}
    \item The length of function code exceeds the max output limit (i.e., 512 tokens in our settings).
    \item The function does not have any modifiers or annotations. This helps us to perform post-processing more easily.
    \item The function is standalone, which means that it can be moved to a new class\footnote{We assume that all Java standard libraries are imported in advance when applying this rule.} and still compiles. The removal of standalone functions makes the benchmark more complex and difficult, and helps better evaluate the performance of repository-level code generation.
\end{enumerate}

We apply the filtering rules sequentially: 17 functions are filtered out by the first rule, 5 by the second, and 17 by the third. As a result, after removing these 39 functions, we obtain 199 coding tasks. As the methods in the benchmark are mined by the CoderUJB authors from the Defects4J~\cite{just2014defects4j} dataset, it is convenient to evaluate the correctness of the generated code using the existing extensive unit test methods. Specifically, we ``patch'' the repository with generated code and run the \code{defects4j test -t ...} command.

\subsubsection{Rust Benchmark}

To our knowledge, the only code generation benchmark for the Rust programming language is MultiPL-E~\cite{cassano2023multipl}, which translates the original HumanEval~\cite{humaneval} and MBPP~\cite{mbpp} benchmarks. However, this benchmark is designed for standalone function code generation tasks and thus is not suitable for our evaluation. In this case, we construct a new benchmark named \textit{RustEval}.

\begin{table}[htbp]
\caption{The crates used in our Rust benchmark}
\label{tab:rust_crates}
\begin{tabular}{cc||cc}
\toprule
\textbf{Crate} & \textbf{\# Downloads} & \textbf{Crate} & \textbf{\# Downloads} \\ \midrule
bit-vec        & 54M                   & sdl2           & 1.7M \\ 
curv-kzen      & 161K                  & splay\_tree    & 72K  \\
dlv-list       & 14M                   & sshkeys        & 189K \\
gptman         & 514K                  & unix\_path     & 86K  \\
mathru         & 202K                  & vec\_map       & 81M  \\
multiset       & 99K                   & without-alloc  & 259K \\
rocket\_http   & 4.2M                  &                &      \\
\bottomrule
\end{tabular}
\end{table}

As shown in Table~\ref{tab:rust_crates}, we collect 13 repositories of various categories from \textit{crates.io}~\cite{cratesio}, the official Rust package registry. In these repositories, we obtain 90 functions with docstrings that satisfy the official guidelines~\cite{rustdoc} (i.e., have detailed descriptions and at least one documentation test). To test the correctness of the generated code, we utilize the documentation tests in the docstrings by running the \code{cargo test --doc ...} command.

\subsubsection{Benchmark Statistics}

\begin{table}[htbp]
\caption{Statistics of our benchmarks, including number of tasks, average number of lines of code (NLOC) per task, and average cyclomatic complexity number (CCN) per task.}
\label{tab:stat}
\begin{tabular}{@{}lcc@{}}
\toprule
                   & \textbf{Java} & \textbf{Rust} \\ \midrule
\textbf{\# Tasks}  & 199           & 90            \\
\textbf{Avg. NLOC} & 16.59         & 10.68         \\
\textbf{Avg. CCN}  & 4.59          & 2.64          \\ \bottomrule
\end{tabular}
\end{table}

Table~\ref{tab:stat} presents the statistics of our benchmarks, where we assess their complexity in terms of the average number of lines of code (NLOC) and cyclomatic complexity number (CCN) per task. Compared to the code completion task in our baseline paper, which requires generating only a single line of code, our function code generation task involves producing more lines and deeper code structures, as reflected in the NLOC and CCN values shown in Table~\ref{tab:stat}. In addition, compared to the Java benchmark, the Rust benchmark is less syntactically verbose and structurally simpler. We attribute this to idiomatic Rust’s reduced boilerplate and more concise abstractions. However, from a semantic standpoint, this does not imply that code generation is easier, as the underlying logic can still be dense and complex.

\subsection{Implementation}
\label{sec:impl}

\subsubsection{Approach}

We implement \appname for Java and Rust, and the source code is written in Python and Rust. Specifically in \textit{Relevant Code Retrieval}, we use MPNet V2~\cite{mpnetv2} as the embedding model. It is an open-source model fine-tuned from the base MPNet~\cite{song2020mpnet} model using massive training data, including the popular CodeSearchNet~\cite{husain2019codesearchnet}. For evaluation on the benchmarks, the ground truth code in the repositories is removed during retrieval to prevent leakage.
In \textit{Type Context Extraction}, we use Eclipse JDT.LS~\cite{eclipsejdtls} and rust-analyzer~\cite{rust_analyzer} as the static analyzers. For Java, our approach interacts with Eclipse JDT.LS through the Language Server Protocol~\cite{lsp_spec}. For Rust, we use rust-analyzer as a third-party library and implement a binding between Rust and Python.

\subsubsection{Model Selection}
\label{sec:model_selection}

\begin{table}[htbp]
\caption{Models used for evaluation in RQ3. The model acronyms correspond to the $x$-axis labels in Figure~\ref{fig:rq3}.}
\label{tab:rq3_list}
\begin{tabular}{@{}cllc@{}}
\toprule
\multicolumn{2}{c}{\textbf{Model ID}}  & \textbf{Acronym} & \textbf{Open-Source?} \\ \midrule
\multicolumn{4}{c}{\cellcolor[HTML]{EFEFEF} \textbf{Code Specialized}}    \\
\cite{codellama}     & \textit{CodeLlama-7b-Instruct}        & CL-7B   & \dingcheckmark \\
\cite{codellama}     & \textit{CodeLlama-13b-Instruct}       & CL-13B  & \dingcheckmark \\
\cite{deepseekcoder} & \textit{deepseek-coder-6.7b-instruct} & DS-6.7B & \dingcheckmark \\
\cite{codegemma}     & \textit{codegemma-7b-it}              & CG-7B   & \dingcheckmark \\
\cite{codeqwen}      & \textit{CodeQwen1.5-7B-Chat}          & CQ-7B   & \dingcheckmark \\
\midrule
\multicolumn{4}{c}{\cellcolor[HTML]{EFEFEF} \textbf{General Purpose}}     \\
\cite{llama3}        & \textit{Meta-Llama-3-8B-Instruct}     & ML-8B   & \dingcheckmark \\
\bottomrule
\end{tabular}
\end{table}

For RQ1 and RQ2, we perform the evaluation using \textit{CodeLlama-13B-Instruct}~\cite{codellama} as the default LLM, due to its state-of-the-art performance among models of medium size. For RQ3, to evaluate the generalizability of \appname, we select a variety of state-of-the-art LLMs with different sizes, including code-specialized models and general-purpose models. The details of the models are listed in Table~\ref{tab:rq3_list}, in which the model IDs are taken from the official releases at HuggingFace\cite{huggingface} or the API documentations.

\subsubsection{Environment and Model Access}

The experiments are performed on a Linux server with capable NVIDIA GPUs. For open-source LLMs, we deploy a local API server based on vLLM~\cite{vllm}, a unified inference and serving engine for LLMs. The weights of open-source models are available at HuggingFace~\cite{huggingface}. To facilitate future research, we list URLs of the models in our replication package.

\subsubsection{Configurations}
\label{sec:exp_config}

In \textit{Relevant Code Retrieval}, we set the maximum chunk size to 2,000 and $k$ (i.e., the number of documents that a retriever should return) to 4 for Java repositories. We set the chunk size to 1,000 and $k$ to 8 for Rust repositories. This is due to the observation that functions and docstrings in our Java repositories are generally longer. In addition, when performing re-ranking, we set the weights of the sparse and dense retrievers to 30\% and 70\%, respectively.

In evaluations, we set $n$ in compile@$k$ and pass@$k$ to 10 and $k$ to 1, 3, and 5. To control the randomness of the generated code~\cite{humaneval}, we set the temperature to 0.6 and the nucleus sampling~\cite{nucleus} parameter (\textit{top\_p}) to 0.7. In addition, we set the maximum number of generated tokens to 512.

\section{Experimental Results}
\label{sec:results}

In this section, we analyze the experiment results in detail and answer the research questions.

\subsection{RQ1: Effectiveness}

\subsubsection{Results}

\begin{table}[t]
\caption{Evaluation results of different methods on two benchmarks, using \textit{CodeLlama-13B-Instruct}}
\label{tab:rq1}
\tabcolsep=4pt
\begin{tabular}{|cllllllllllll|}
\hline
\multicolumn{1}{|c|}{} &
  \multicolumn{6}{c|}{\textbf{Java}} &
  \multicolumn{6}{c|}{\textbf{Rust}} \\ \cline{2-13} 
\multicolumn{1}{|c|}{} &
  \multicolumn{3}{c|}{\textit{compile@k} (\%)} &
  \multicolumn{3}{c|}{\textit{pass@k} (\%)} &
  \multicolumn{3}{c|}{\textit{compile@k} (\%)} &
  \multicolumn{3}{c|}{\textit{pass@k} (\%)} \\ \cline{2-13} 
\multicolumn{1}{|c|}{\multirow{-3}{*}{\textbf{Method}}} &
  \multicolumn{1}{c}{$1$} &
  \multicolumn{1}{c}{$3$} &
  \multicolumn{1}{c|}{$5$} &
  \multicolumn{1}{c}{$1$} &
  \multicolumn{1}{c}{$3$} &
  \multicolumn{1}{c|}{$5$} &
  \multicolumn{1}{c}{$1$} &
  \multicolumn{1}{c}{$3$} &
  \multicolumn{1}{c|}{$5$} &
  \multicolumn{1}{c}{$1$} &
  \multicolumn{1}{c}{$3$} &
  \multicolumn{1}{c|}{$5$} \\ \hline
\multicolumn{1}{|c|}{\vanilla} &
  33.6 &
  49.4 &
  \multicolumn{1}{l|}{56.2} &
  14.9 &
  22.1 &
  \multicolumn{1}{l|}{25.2} &
  18.0 &
  28.3 &
  \multicolumn{1}{l|}{33.1} &
  10.8 &
  15.8 &
  18.2 \\
\multicolumn{1}{|c|}{\infile} &
  59.7 &
  72.7 &
  \multicolumn{1}{l|}{76.8} &
  35.3 &
  43.8 &
  \multicolumn{1}{l|}{47.6} &
  52.1 &
  66.5 &
  \multicolumn{1}{l|}{70.9} &
  41.6 &
  55.7 &
  61.6 \\
\multicolumn{1}{|c|}{\repocoder} &
  63.0 &
  74.1 &
  \multicolumn{1}{l|}{78.2} &
  41.0 &
  46.9 &
  \multicolumn{1}{l|}{49.0} &
  63.7 &
  73.4 &
  \multicolumn{1}{l|}{76.4} &
  49.4 &
  59.5 &
  63.0 \\
\rowcolor[HTML]{ECF4FF} 
\multicolumn{1}{|c|}{\cellcolor[HTML]{ECF4FF}\textbf{\appname}} &
  \textbf{71.9} &
  \textbf{84.8} &
  \multicolumn{1}{l|}{\cellcolor[HTML]{ECF4FF}\textbf{88.4}} &
  \textbf{44.7} &
  \textbf{54.0} &
  \multicolumn{1}{l|}{\cellcolor[HTML]{ECF4FF}\textbf{57.5}} &
  \textbf{65.1} &
  \textbf{75.5} &
  \multicolumn{1}{l|}{\cellcolor[HTML]{ECF4FF}\textbf{78.9}} &
  \textbf{52.7} &
  \textbf{62.0} &
  \textbf{65.9} \\ \hline
\end{tabular}
\end{table}

The overall result is presented in Table~\ref{tab:rq1}. According to Table~\ref{tab:rq1}, we observe that the \vanilla baseline performs poorly as expected because the lack of repository context is likely to cause hallucinations. Next, we can conclude that the \infile baseline performs worse than \repocoder and \appname, as in-file context is not enough to provide key information such as cross-file type structures and API usage. In addition, \appname outperforms \repocoder. In terms of the Java benchmark, the compile@$k$ and pass@$k$ scores increase by up to 14.44\% and 17.35\%, respectively. For the Rust benchmark, the compile@$k$ and pass@$k$ scores increase by up to 3.27\% and 6.68\%. 

\begin{figure}[t]
    \centering
    \includegraphics[width=\linewidth]{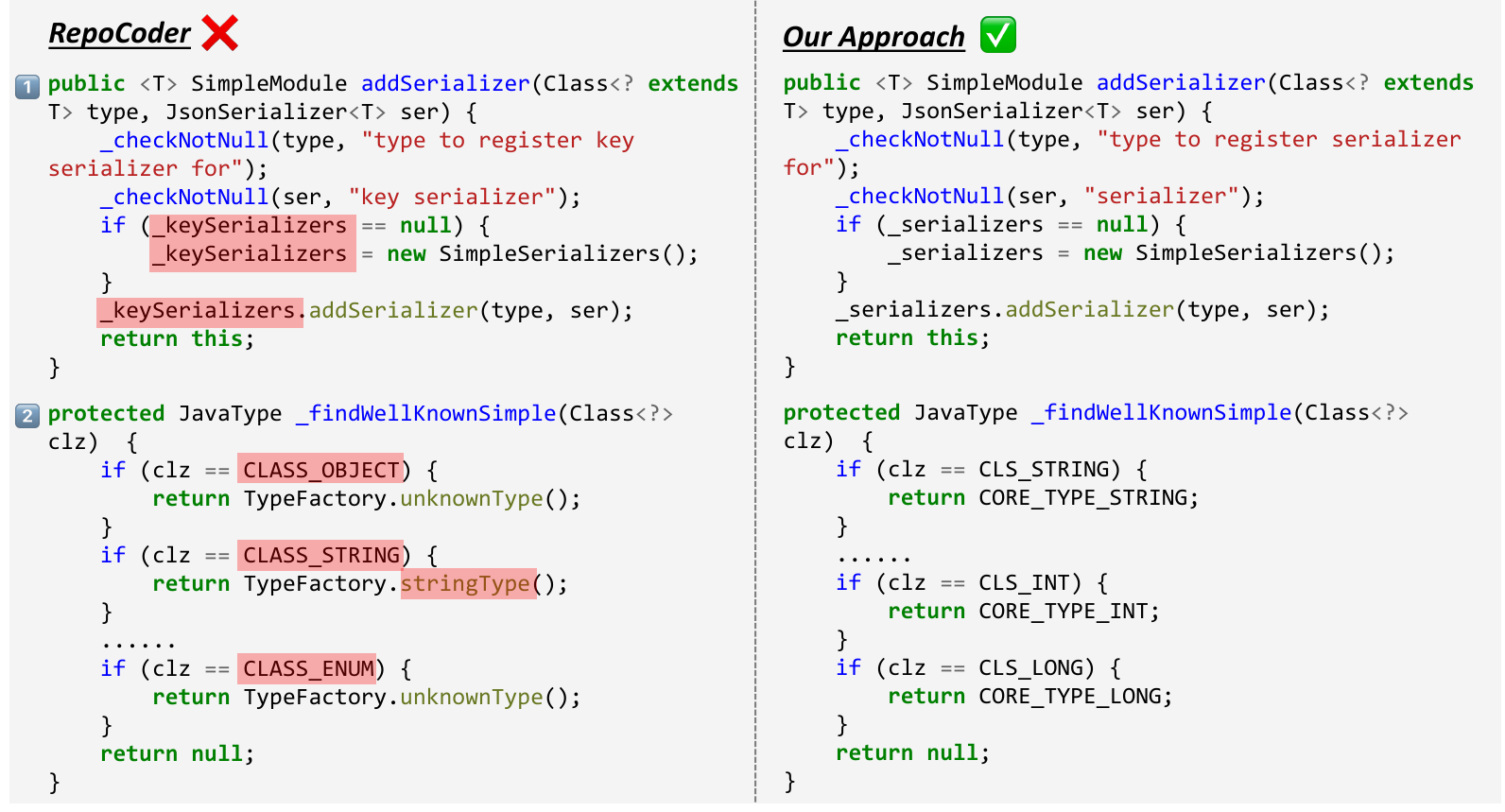}
    \caption{Two tasks in the Java benchmark, where \repocoder fails but \appname succeeds.}
    \label{fig:rq1_case}
\end{figure}

\subsubsection{Qualitative Analysis}
\label{sec:rq1_qual}

To show the advantages of \appname over \repocoder, we also perform a qualitative analysis of the results. Fig.~\ref{fig:rq1_case} presents two cases, in which \repocoder fails to generate the correct function code while \appname succeeds.

\myding{1} \textbf{Case 1} (method \code{addSerializer}): The code generated by \repocoder compiles but fails to pass unit tests because it uses the wrong field \code{_keySerializers} instead of the correct one (i.e., \code{_serializers}). 

Upon checking the context provided to the LLM, we find that \repocoder retrieves a similar method named \code{addKeySerializers}, which uses the field \code{_keySerializers} in its body. The LLM simply copies the code from \code{addKeySerializer} and causes the code to fail in unit tests. For \appname, it also retrieves the method \code{addKeySerializer}, and the pattern of its generated code is the same as the code generated by \repocoder. However, with the type context extracted by \appname, the LLM knows the existence of pairs (field \code{_serializers}, method \code{addSerializer}) and (field \code{_keySerializers}, method \code{addKeySerializers}). Therefore, we believe that this makes the LLM filter out some misleading reference code and choose the right field for usage.

\myding{2} \textbf{Case 2} (method \code{_findWellKnownSimple}): The code generated by \repocoder causes many compilation errors, as it contains non-existent fields and methods in the current scope. 

In detail, we find that the non-existent \code{CLASS_XXX} fields come from another file, and \repocoder retrieves them as context. Without knowing the type context of \code{JavaType} and current class (i.e., \code{TypeFactory}), the LLM accepts misleading references and generates a non-existent API call \code{stringType()}.

\subsubsection{Comparison of Java and Rust}

As shown in Table~\ref{tab:rq1}, the magnitude of improvement provided by \textsc{CatCoder} over the \textsc{RepoCoder} baseline differs between the two languages. On the Java benchmark, \textsc{CatCoder} achieves a relative improvement of up to 17.35\% in pass@k, whereas on Rust, the maximum relative improvement is 6.68\%. This suggests that \textsc{CatCoder} offers a more substantial advantage in the context of the evaluated Java projects. We attribute this difference to the characteristics of the languages and the structural properties of the benchmark projects. As indicated in Table~\ref{tab:stat}, the Java benchmark is, on average, more structurally complex and verbose than the Rust benchmark. We argue that this increased complexity creates scenarios in which the type context provided by \textsc{CatCoder} becomes a more critical factor for successful code generation, as illustrated in Section~\ref{sec:rq1_qual}.

\begin{table}[t]
\caption{Results of CodeBLEU, using \textit{CodeLlama-13B-Instruct}. \textbf{C}: CodeBLEU score; \textbf{N}: n-gram match score; \textbf{W}: weighted n-gram match score; \textbf{S}: syntax match score; \textbf{D}: dataflow match score.}
\label{tab:semantic}
\setlength{\tabcolsep}{4pt}
\begin{tabular}{|c|ccccc|cccc|}
\hline
                                  & \multicolumn{5}{c|}{\textbf{Java}}                             & \multicolumn{4}{c|}{\textbf{Rust}}                \\ \cline{2-10} 
\multirow{-2}{*}{\textbf{Method}} & \textit{C} & \textit{N} & \textit{W} & \textit{S} & \textit{D} & \textit{C} & \textit{N} & \textit{W} & \textit{S} \\ \hline
\vanilla                          & 0.36       & 0.23       & 0.30       & 0.41       & 0.37       & 0.25       & 0.16       & 0.21       & 0.39       \\
\infile                           & 0.51       & 0.40       & 0.46       & 0.58       & 0.55       & 0.38       & 0.29       & 0.33       & 0.54       \\
\repocoder                        & 0.57       & 0.47       & 0.53       & 0.64       & 0.61       & 0.46       & 0.37       & 0.42       & 0.60       \\
\rowcolor[HTML]{ECF4FF} 
\appname                          & 0.56       & 0.46       & 0.52       & 0.63       & 0.59       & 0.44       & 0.35       & 0.40       & 0.60       \\ \hline
\end{tabular}
\end{table}

\subsubsection{Semantic Similarity vs. Test Pass Rate}

In addition to compile@k and pass@k, we conduct an evaluation using CodeBLEU~\cite{ren2020codebleu}, a metric that lies between purely text similarity-based and execution-based metrics. Beyond traditional n-gram matching, CodeBLEU incorporates syntactic and semantic matching, thus better capturing the semantic similarity between code snippets. It is computed as: 
$$\mathrm{CodeBLEU} = \alpha \cdot \mathrm{BLEU} + \beta \cdot \mathrm{BLEU_{weight}} + \gamma \cdot \mathrm{Match_{ast}} + \delta \cdot \mathrm{Match_{df}},$$
where $\mathrm{BLEU}$ is the n-gram match score, $\mathrm{BLEU_{weight}}$ is the weighted n-gram match score, $\mathrm{Match_{ast}}$ is the abstract syntax tree (AST) match score, and $\mathrm{Match_{df}}$ is the dataflow match score. The parameters $\alpha, \beta, \gamma, \text{and } \delta$ are weighting coefficients.

We use an open-source implementation\footnote{\url{https://github.com/k4black/codebleu}} of CodeBLEU, and the results are presented in Table~\ref{tab:semantic}. For the Java benchmark, the weights are set to $\alpha=\beta=\gamma=\delta=0.25$. For the Rust benchmark, since the tool cannot extract dataflow from Rust code, we omit the dataflow match component and assign the weights to $\alpha=\beta=\gamma=0.33$ and $\delta=0$.

On both Java and Rust benchmarks, \textsc{CatCoder} and \textsc{RepoCoder} outperform the other two baselines, \textsc{Vanilla} and \textsc{In-File}, across CodeBLEU and all four sub-metrics, suggesting that relevant code retrieval helps LLMs generate code more similar to the ground truth. Notably, the scores of \textsc{CatCoder} and \textsc{RepoCoder} are very close, with \textsc{RepoCoder} even slightly outperforming \textsc{CatCoder} in CodeBLEU. However, since \textsc{CatCoder} achieves higher compile@k and pass@k scores, particularly on the Java benchmark, this indicates that \textsc{RepoCoder} sometimes produces code structurally similar to the ground truth but fails due to subtle issues. This underscores the importance of execution-based metrics in evaluating repository-level code generation.

\rqans{1}{\appname is effective at generating repository-level function code, and it outperforms the \repocoder baseline by up to 14.44\% and 17.35\% in terms of compile@$k$ and pass@$k$ scores. Besides, in qualitative analysis, we find that the type context is effective in filtering out misleading reference code.}

\subsection{RQ2: Ablation Study}

\begin{table}[t]
\caption{Evaluation results of different variants on two benchmarks, using \textit{CodeLlama-13B-Instruct}}
\label{tab:rq2}
\tabcolsep=4pt
\begin{tabular}{|c|llllll|llllll|}
\hline
\multirow{3}{*}{\textbf{Variant}} & \multicolumn{6}{c|}{\textbf{Java}}   & \multicolumn{6}{c|}{\textbf{Rust}}   \\ \cline{2-13} 
 &
  \multicolumn{3}{c|}{\textit{compile@k} (\%)} &
  \multicolumn{3}{c|}{\textit{pass@k} (\%)} &
  \multicolumn{3}{c|}{\textit{compile@k} (\%)} &
  \multicolumn{3}{c|}{\textit{pass@k} (\%)} \\ \cline{2-13} 
 &
  \multicolumn{1}{c}{$1$} &
  \multicolumn{1}{c}{$3$} &
  \multicolumn{1}{c|}{$5$} &
  \multicolumn{1}{c}{$1$} &
  \multicolumn{1}{c}{$3$} &
  \multicolumn{1}{c|}{$5$} &
  \multicolumn{1}{c}{$1$} &
  \multicolumn{1}{c}{$3$} &
  \multicolumn{1}{c|}{$5$} &
  \multicolumn{1}{c}{$1$} &
  \multicolumn{1}{c}{$3$} &
  \multicolumn{1}{c|}{$5$} \\ \hline
\rowcolor[HTML]{ECF4FF} 
FULL         & 71.9 & 84.8 & \multicolumn{1}{l|}{88.3} & 44.7 & 54.0 & 57.5 & 65.1 & 75.5 & \multicolumn{1}{l|}{78.9} & 52.7 & 62.0 & 65.9 \\
-CR          & 57.9 & 75.4 & \multicolumn{1}{l|}{81.4} & 27.5 & 38.8 & 43.7 & 35.0 & 50.7 & \multicolumn{1}{l|}{56.3} & 23.3 & 35.4 & 41.0 \\
-TC          & 67.1 & 80.3 & \multicolumn{1}{l|}{84.8} & 39.5 & 49.5 & 53.4 & 59.6 & 70.1 & \multicolumn{1}{l|}{73.8} & 49.9 & 59.6 & 63.8 \\ \hline
\end{tabular}
\end{table}

To investigate the contribution of the components (as described in Section~\ref{sec:method}) to the effectiveness of
\appname, we create the following variants:

\begin{enumerate}
    \item FULL: The complete version of \appname, exactly as Section~\ref{sec:method} illustrates.
    \item -CR: \textit{Relevant Code Retrieval} is removed.
    \item -TC: \textit{Type Context Extraction} is removed.
\end{enumerate}

We evaluate the above three variants on the same benchmarks as RQ1, also using \textit{CodeLlama-13B-Instruct} as the underlying LLM. We compare the incomplete ones with FULL, and the results are illustrated in Table~\ref{tab:rq2}. We can conclude that \appname benefits from the two components, as the compile@$k$ and pass@$k$ scores of both incomplete variants drop noticeably. In detail, for the -CR variant, the compile@$k$ and pass@$k$ scores decrease by up to 19.43\% and 38.35\%, respectively, on the Java benchmark, and by up to 46.24\% and 56.03\% on the Rust benchmark. And for the -TC variant, the compile@$k$ and pass@$k$ scores drop by up to 6.63\% and 11.57\% respectively on the Java benchmark, and by up to 8.52\% and 5.3\% on the Rust benchmark.

Notably, removing \textit{Type Context Extraction} leads to a smaller performance decline compared to removing \textit{Relevant Code Retrieval}.
This observation underscores the effectiveness of our hybrid retrieval mechanism. However, we emphasize that the importance of type context is not diminished. In many instances, particularly those involving functionalities that are symmetric to, or minor variations of, existing code in the repository, the retriever is able to find highly relevant examples. The presence of duplicated code snippets benefits retrieval-based generation by providing the LLM with strong, nearly complete templates for the target function. In such cases, the retrieved code is sufficient for the LLM to produce a correct solution, rendering the additional type context less critical. However, the performance gap between the -TC variant and FULL highlights a subset of more challenging problems where retrieval alone is insufficient. The qualitative examples in Fig.~\ref{fig:rq1_case} illustrate a typical failure mode: the retrieved code may include distracting snippets or reference elements outside the current scope, resulting in generation errors. Therefore, we argue that our Type Context Extraction component is an essential and complementary mechanism.

\rqans{2}{The two components of \appname, namely \textit{Relevant Code Retrieval} and \textit{Type Context Extraction}, improve the effectiveness of \appname.}

\subsection{RQ3: Generalizability}

\begin{figure}
    \centering
    \includegraphics[width=\linewidth]{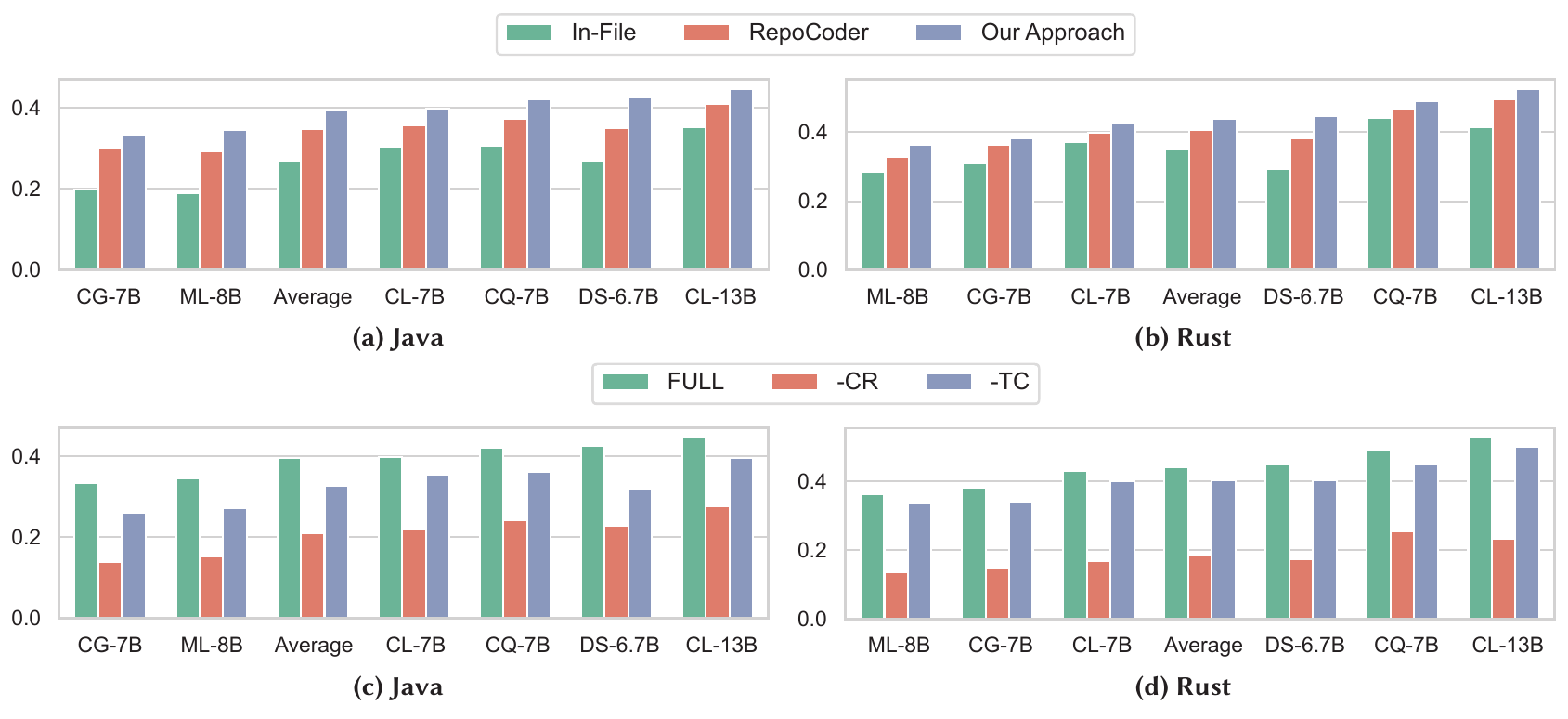}
    \caption{\textbf{pass@1} performance comparison using various LLMs on the Java and Rust benchmarks. Subfigures (a) and (b) compare \appname with baseline methods—\textsc{In-File} and \textsc{RepoCoder}—while (c) and (d) compare \appname with its variants. Model acronyms on the $x$-axis are consistent with those listed in Table~\ref{tab:rq3_list}, and variant names refer to those introduced in RQ2. Models are sorted in ascending order based on the performance of \appname. The “Average” bar represents the mean pass@1 score across all evaluated models.}
    \label{fig:rq3}
\end{figure}

To assess whether the benefits of \textsc{CatCoder} are specific to a single LLM or generalize across different models, we conduct an extensive generalizability study. As described in Section~\ref{sec:model_selection}, we select a diverse set of LLMs, including both code-specialized and general-purpose models of varying sizes (Table~\ref{tab:rq3_list}). We then evaluate the performance of the \textsc{In-File}, \textsc{RepoCoder}, and \textsc{CatCoder} approaches, as well as the \textsc{CatCoder} variants (-CR and -TC), using each of these models on both our Java and Rust benchmarks.

\subsubsection{Results on Generalizability}

Fig.~\ref{fig:rq3} shows the pass@$1$ scores for these experiments. 
The results for both Java (Fig.~\ref{fig:rq3}a,~\ref{fig:rq3}c) and Rust (Fig.~\ref{fig:rq3}b,~\ref{fig:rq3}d) benchmarks suggest that \appname is able to improve the performance of every tested model. 

On the Java benchmark, \appname improves the average pass@$1$ score by 13.73\%, compared to \repocoder (Fig.~\ref{fig:rq3}a). The improvement is universal, ranging from a significant increase for models like \textit{DS-6.7B} to a more modest increase for models like \textit{CL-13B}. Similarly, on the Rust benchmark (Fig.~\ref{fig:rq3}b), \appname consistently outperforms both \textsc{In-File} and \textsc{RepoCoder} for all LLMs. It improves the average pass@$1$ score by 8.4\% compared to \textsc{RepoCoder}. Additionally, the relative performance ranking of the models remains largely consistent across both languages, with \textit{CL-13B} and \textit{DS-6.7B} generally at the top, and with \textit{CG-7B} and \textit{ML-8B} showing more modest results.

Furthermore, the ablation study results, when generalized across models (Fig.~\ref{fig:rq3}c,~\ref{fig:rq3}d), are consistent with the findings from RQ2.
For both Java and Rust, the -CR variant (without Code Retrieval) consistently leads to the most significant performance degradation, while the -TC variant (without Type Context) shows a smaller but still noticeable drop.
On average, across all models on the Java benchmark, the pass@$1$ score decreases by 46.89\% without \textit{Relevant Code Retrieval}, and decreases by 17.17\% without \textit{Type Context Extraction}.
This pattern holds for Rust as well. On average, removing \textit{Relevant Code Retrieval} decreases the pass@1 score by 57.94\%, while removing \textit{Type Context Extraction} leads to an 8.21\% decrease.

In conclusion, our results support the generalizability of \appname. It consistently enhances the performance of a wide variety of LLMs on both Java and Rust, demonstrating that the architectural benefits of \appname are model-agnostic.

\subsubsection{Comparison of different LLMs}

The results of our generalizability study also offer insights into the current landscape of LLMs for code:

\myding{1} Although the selected general purpose LLM (i.e., \textit{ML-8B}) claims to be powerful in many aspects~\cite{llama3}, it falls behind all selected code-specialized models except for \textit{CG-7B} in our benchmarks. One example is \textit{ML-8B} versus \textit{CL-7B}. Although \textit{CL-7B} is fine-tuned from an older foundation model (i.e., Llama 2~\cite{touvron2023llama2}), it outperforms the new model \textit{ML-8B} (i.e., Llama 3~\cite{llama3}). This indicates that code generation tasks benefit a lot from fine-tuning models using high-quality code datasets.

\myding{2} From the trend in Fig.~\ref{fig:rq3}, although larger models tend to have a better absolute performance than smaller models, the improvement by \appname is larger on smaller models. For instance, the pass@1 score increase for models like \textit{ML-8B}, \textit{CQ-7B}, and \textit{DS-6.7B} is more pronounced than the increase for \textit{CL-13B}. This suggests that providing high-quality, structured context via frameworks like \appname can help smaller or less specialized models bridge the performance gap with larger, more powerful ones.

\rqans{3}{\appname is generalizable to many different LLMs, as it improves the pass@$1$ scores of all selected LLMs, compared with both \infile method and \repocoder. In addition, the performance of different LLMs on the benchmark provides valuable insights.}

\subsection{RQ4: Scalability}

\begin{figure}[htbp]
    \centering
    \includegraphics[width=0.6\linewidth]{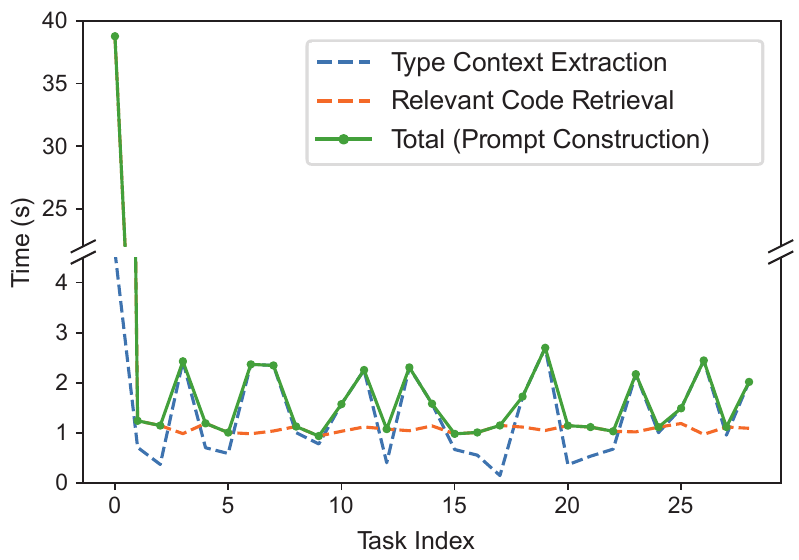}
    \caption{Detailed latency of the components when constructing prompts for 29 code generation tasks sequentially in the \textit{closure-compiler} project.}
    \label{fig:efficiency}
\end{figure}

\begin{table}[htbp]
    \centering
    \caption{Average latency of the components when constructing prompts for 29 code generation tasks sequentially in the \texttt{closure-compiler} project. CR: Relevant Code Retrieval, in Section~\ref{sec:retrieval}; TC: Type Context Extraction, in Section~\ref{sec:context}.}
    \label{tab:efficiency}
    \begin{tabular}{@{}cccc@{}}
    \toprule
    \textbf{Case}            & \textbf{CR} & \textbf{TC} & \multicolumn{1}{l}{\textbf{Total}} \\ \midrule
    Cold start (Task 0)      & 38.8s       & 4.62s         & 38.8s      \\
    Avg. after cold start    & 1.08s       & 1.32s         & 1.56s      \\
    \midrule
    Avg. speedup w/ cache & 36x         & 3.5x          & 25x        \\
    \bottomrule
    \end{tabular}
\end{table}

In real-world software development, the latency of repository-level code generation tools is crucial, as they inherently process considerable volumes of contextual code in large and complex projects. Significant latencies would negate the advantages of automated assistance and lead to user frustration. Therefore, we propose this research question and investigate the scalability of \appname when applied to functions within large-scale code repositories.

\subsubsection{Experimental Design}

To robustly assess \appname's scalability, the \textit{closure-compiler} repository\footnote{The Closure Compiler is a tool for checking and optimizing JavaScript code. The project is developed by Google and has attracted wide attention.} in our benchmark (Section~\ref{sec:benchmarks}) is selected for evaluation, by considering the following aspects:

\begin{itemize}
    \item This repository is the largest one in our benchmark and comprises approximately 400,000 lines of code, representing a non-trivial, real-world system. Thus, it can present a genuine scalability challenge for \appname.
    \item It is a well-known and actively developed open-source project. Thus, the evaluation is realistic and can reflect the types of complex dependencies, code structures, and API usage patterns that \appname would encounter in practical deployment.
\end{itemize}

Specifically, our benchmark contains 29 distinct function code generation tasks for the \textit{closure-compiler} repository. For evaluation, we open this repository, invoke \appname on these code generation tasks sequentially, and examine the time consumption of each code generation task. This sequential execution simulates a common developer workflow where a programmer might work on several functions or address multiple issues within the same project during a single coding session.

For time consumption, we focus on the latency of \appname's core components (i.e., Relevant Code Retrieval and Type Context Extraction), while excluding the inference latency of the underlying LLM. This is due to that the time taken for LLM inference is highly dependent on external factors that are largely independent of \appname, including the specific LLM chosen, the underlying hardware, and the deployment environment. In addition, the length of prompts constructed by \appname does not scale with repository sizes, making the inference latency less important in scalability evaluation.

\subsubsection{Analysis of Scalability}

Fig.~\ref{fig:efficiency} shows the detailed latency of the two components of \appname when running each sequential code generation task in the \textit{closure-compiler} repository. In addition, Table~\ref{tab:efficiency} shows the statistics of average latency. It is also worth noting that the two components are independent and can run in parallel. Thus, the total time is bounded by the slower component.

We can notice that \appname takes a longer time to process the very first code generation task in a repository that it has not previously analyzed. We refer to the first task as ``cold start''. The latency of cold start can be attributed to several setup operations in our approach:

\begin{itemize}
    \item For Relevant Code Retrieval (CR), the setup operations mainly involve generating embedding vectors for code chunks, which are crucial for subsequent semantic search.
    \item For Type Context Extraction (TC), the operations include the static analyzer's setup processes, such as parsing the codebase and building representations of type hierarchies and dependencies.
\end{itemize}

As shown in Table~\ref{tab:efficiency}, the measured cold start time is 38.8 seconds, with the Relevant Code Retrieval component as the main bottleneck. Considering that the \textit{closure-compiler} repository is large, and it also takes many IDEs to index a new large repository for a long time, we believe that the cold start overhead of \appname is still acceptable for enabling its advanced context-aware code generation capabilities.

Following the initial cold start, \appname's performance for subsequent tasks improves significantly. Table~\ref{tab:efficiency} shows the reduction in average latency for tasks after the first one: Relevant Code Retrieval time drops to 1.08s, Type Context Extraction to 1.32s, and the total prompt construction time averages 1.56s. The results indicate an average 36x speedup for Relevant Code Retrieval, 3.5x speedup for Type Context Extraction, and 25x speedup for total time. In this case, after the initial waiting, users do not have to wait for a long time each time to receive a response in an interactive development workflow. This is due to \appname's caching mechanism, which stores intermediate data, including embedding vectors of code chunks and intermediate states of the static analyzer. The cached data is computationally expensive to generate, but generally stable across a sequence of related development tasks within a single coding session. The massive speedup after the cold start demonstrates the effective reuse of cache data.

\rqans{4}{\appname shows practical scalability at a series of code generation tasks within large repositories. In the case of the \textit{closure-compiler} repository, the caching mechanism of \appname is able to introduce a 25x average speedup after the cold start.}

\section{Discussion}
\label{sec:discussion}

In this section, we discuss the limitations of our approach and outline potential directions for future work, followed by a discussion of threats to validity.

\subsection{Limitations of \textsc{CatCoder}}

\subsubsection{Limitation in the Underlying LLM}

\begin{figure}[htbp]
    \centering
    \includegraphics[width=\linewidth]{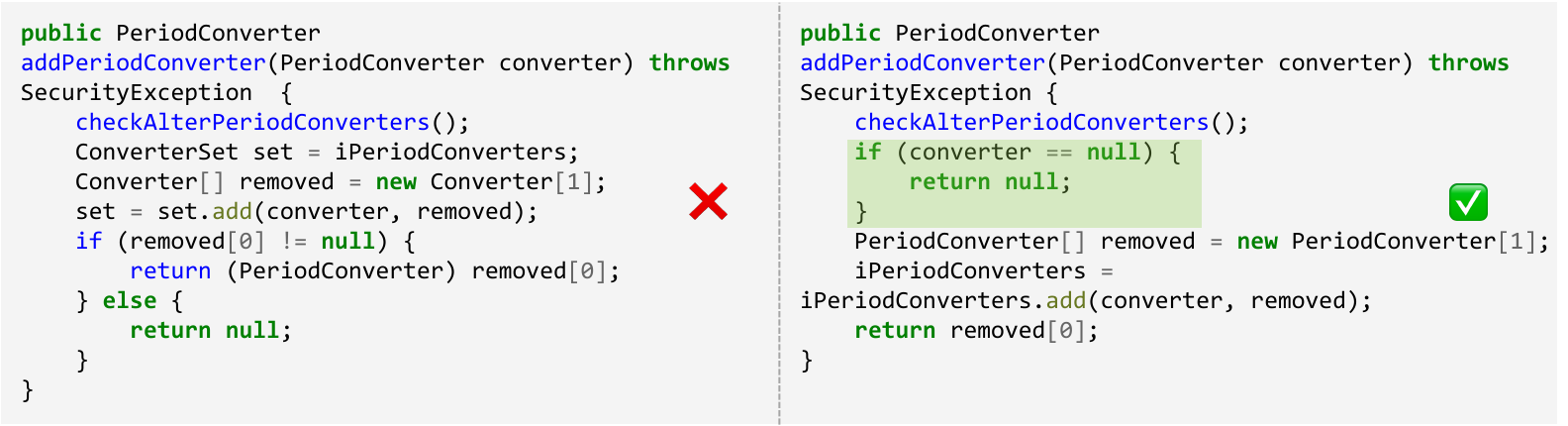}
    \caption{A case that \appname fails to generate correct code. The first code is generated by \appname, and the second one is the ground truth.}
    \label{fig:fail_case}
\end{figure}

Apart from relevant code and type context, an LLM's internal capability of understanding general syntax and algorithmic principles is equally essential for correct repository-level code generation.
Similar to many prior studies~\cite{liu2023repobench,wu2024repoformer,zhang2023repocoder,shrivastava2023repository}, \appname operates on a frozen LLM. As such, this line of work is inherently limited in enhancing the model's logical reasoning abilities.

Fig.~\ref{fig:fail_case} presents a failure case produced by \appname. Although the generated code correctly uses all relevant APIs and closely resembles the ground truth implementation, it fails to pass some unit tests due to missing the logic for handling \verb|null| pointers (highlighted in green in Fig.~\ref{fig:fail_case}). This example illustrates that function code generation is inherently complex, and the effective use of repository context does not always guarantee correctness. A possible direction for future work could be fine-tuning LLMs on code repositories to enhance repository-level in-context learning, or adopting reinforcement learning approaches~\cite{le2022coderl} to improve reasoning and coding capabilities.

\subsubsection{Limitation in Retrieval}

In the Relevant Code Retrieval component, we adopt a hybrid retrieval strategy that combines sparse (BM25) and dense (embedding-based) methods to achieve high recall and precision. However, like any information retrieval system, it is not infallible. The retriever may still return incomplete or even misleading context. To some extent, the type context in \textsc{CatCoder} helps mitigate this limitation by supplying accessible API information that the LLM can cross-reference, as illustrated in Section~\ref{sec:rq1_qual}. Possible future work includes enhancing the retrieval and reranking pipeline using powerful LLMs. 

\subsubsection{Limitation in Static Analysis}

Although Java is a statically typed programming language, it supports dynamic features such as reflection. The Java Reflection API enables dynamic class loading, object instantiation, and method invocation based on runtime strings. Despite the capabilities of the static analyzer we employ (i.e., Eclipse JDT.LS, commonly used in IDEs), handling these dynamic features remains challenging. For example, if a target function is related to a field whose class is loaded reflectively at runtime (e.g., via the \code{Class.forName} method), the analyzer may fail to resolve the actual class and its members. In such cases, the resulting type context may omit information related to the truly dependent class. Therefore, a limitation is that \textsc{CatCoder} may produce \textit{incomplete} type context when the repository heavily relies on dynamic features.

\subsection{Threats to Validity}

\subsubsection{Internal Threats}

In general, static analyzers may introduce \textit{unsound} results, which can, in turn, affect dependent approaches. However, we argue that this is not a concern in \appname. Specifically, we use static analysis only to locate types and obtain their definitions and implementations. This form of analysis is lightweight and significantly more accurate than other forms, such as pointer or reflection analysis, as it operates solely on explicit type names in the code. Besides, the tools we use (i.e., Eclipse JDT and rust-analyzer) and the analysis we apply have been integrated into widely used IDEs and code editors (e.g., Eclipse and Visual Studio Code). Therefore, we consider further verification of the analysis results in \appname unnecessary, and we trust that our evaluation results remain valid.

\subsubsection{External Threats}

Our implementation of \appname is limited as it only supports Java and Rust. However, the proposed approach does not rely on detailed language specifications and is applicable to any statically typed programming language. In the future, to make \appname more practical, we plan to extend it to support more statically typed languages, for example, C++ and C\#.

\section{Related Work}
\label{sec:related}

\subsection{Large Language Models for Code}

Large language models for code (i.e., code LLMs) refer to a family of Transformer~\cite{vaswani2017attention} models pre-trained or fine-tuned on massive code datasets. Previous work has shown the exceptional performance of code LLMs in various code-related tasks, such as code generation~\cite{humaneval}, test case generation~\cite{tufano2020unit}, and program repair~\cite{program_repair}.
In general, pre-trained language models for code can be categorized into the following three types: 

\begin{enumerate}
    \item \textbf{Encoder-only models}, such as CodeBERT~\cite{feng2020codebert} and GraphCodeBERT~\cite{guo2020graphcodebert}. These models are pre-trained using Masked Language Modeling (MLM) techniques, and are popular and effective for tasks like classification, where a deep understanding of input data is required.
    \item \textbf{Encoder-Decoder models}, such as the CodeT5 family~\cite{wang2021codet5, wang2023codet5+}. These models first encode the input into an internal representation and then decode it into an output sequence.
    \item \textbf{Decoder-only models}, such as Codex~\cite{humaneval}, CodeGen~\cite{nijkamp2022codegen}, StarCoder~\cite{li2023starcoder} and CodeLlama~\cite{codellama}. These models are pre-trained using Next Token Prediction tasks and are powerful in generative tasks.
\end{enumerate}

In our experiments, we select a variety of state-of-the-art code LLMs, all of which are decoder-only.

\subsection{Benchmarks for Repository-Level Code Generation}

Iyer et al.~\cite{concode} are among the first to highlight the importance of the programmatic context and introduce the task of generating class member functions with natural language descriptions and class contexts. In summary, they construct Concode, a large training dataset consisting of Java classes from online code repositories, and propose an encoder-decoder model for code generation. 
With the development of large language models, many subsequent repository-level code generation benchmarks have arisen in academia.
Du et al.~\cite{du2023classeval} make an attempt to evaluate LLMs in the more challenging class-level code generation scenario and manually construct a benchmark named ClassEval, which consists of 100 class-level Python code generation tasks.

Compared with previous benchmarks, for example HumanEval~\cite{humaneval} and MBPP~\cite{mbpp}, ClassEval is more pragmatic. However, it is still limited because, although the methods in the tasks are not standalone, the programmatic contexts are still restricted to very common libraries and methods within the same file. In real-world scenarios, a code repository generally spans a large number of files, making cross-file contexts essential for LLMs to generate the correct code~\cite{ding2024crosscodeeval}.

Toward more realistic evaluation, subsequent benchmarks such as CoderEval~\cite{yu2024codereval}, CrossCodeEval~\cite{ding2024crosscodeeval}, and CoderUJB~\cite{zeng2024coderujb} are proposed, all of which are built on a set of real-world open-source repositories and utilize cross-file context for evaluation. A key novelty of CoderEval and CrossCodeEval lies in their ability to support multilingual code generation tasks, while CoderUJB innovatively spans five kinds of practical programming tasks.

As LLMs evolve, Liu et al.~\cite{liu2023repobench} point out data leakage and memorization problems, which may affect the integrity and trustworthiness of model evaluation results. To mitigate this issue, they propose RepoBench and regularly update the dataset to keep code samples up-to-date. Similarly, Li et al.~\cite{li2024evocodebench} propose EvoCodeBench, which includes an automatic pipeline to evolve itself.

\subsection{Repository-Level Code Generation with LLMs}

Many existing approaches for repository-level code generation perform retrieval. 
Some work, such as~\cite{private_library,lu2022reacc}, searches for useful information from documentation or external databases.
Specifically, Zan et al.~\cite{private_library} train a dual-encoder model, which encodes the natural language descriptions and API information, respectively, to retrieve possible APIs from the documentation of the library.
Lu et al.~\cite{lu2022reacc} propose a retrieval-augmented generation approach named ReAcc, which retrieves similar code from a code database they built. However, these approaches are not flexible and general as their retrieval sources (e.g., API documentation and external databases) may not be available in other repositories.

Other retrieval-based approaches explore retrieving similar code snippets within the repositories.
Shrivastava et al.~\cite{shrivastava2023repository} propose a
framework named Repo-Level Prompt Generator to complete the code for holes in single lines. In the framework, they train a prompt proposal classifier that takes the repository files and a set of prompt proposals and outputs the predicted prompt proposal.
Zhang et al.~\cite{zhang2023repocoder} propose RepoCoder, which utilizes repository information by iteratively retrieving similar code. In the initial iteration, it uses the natural language descriptions as query input, and in subsequent iterations, it uses previously generated code for queries.

Our work \appname combines retrieval-based techniques with a new kind of information called type context, and the comparison with RepoCoder in Section~\ref{sec:results} shows that type context serves as essential auxiliary information for statically typed languages. In addition, \appname does not require model training or multiple generation iterations, making it more practical in real-world settings.

In addition to the above work, which primarily focuses on prompt engineering, several model training frameworks~\cite{wu2024repoformer, deepseekcoder} and decoding strategies~\cite{agrawal2024monitor} have been proposed to support repository-level code generation. As our approach \appname does not modify the underlying LLM, it can be combined with these frameworks to improve effectiveness.

\section{Conclusion and Future Work}
\label{sec:conclusion}

In this paper, we aim to address the lack of type dependency information in repository-level code generation for statically typed programming languages. 
We propose \appname, a code generation framework for statically typed programming languages, which improves the performance of LLMs by retrieving relevant code and extracting type context using static analyzers. 
\appname is systematically evaluated on Java and Rust benchmarks that are built from real-world open-source repositories. 
The results show that \appname outperforms the \repocoder baseline by up to 17.35\% in terms of pass@k scores. 
Further studies demonstrate the generalizability and scalability of \appname. 
We release the replication package of \appname, including the source code of \appname, the benchmark datasets, and the evaluation scripts.
In future work, possible research attempts include investigating the data leakage problem of the benchmarks. These investigations may help further improve the applicability of our proposed approach in practical scenarios.

\begin{acks}
This research is supported by the Fundamental Research Funds for the Central Universities (No. 226-2025-00171).
\end{acks}

\bibliographystyle{ACM-Reference-Format}
\bibliography{ref}

\end{document}